\documentclass[9pt,conference,pdfusetitle]{IEEEtran}

\usepackage[preprint]{waspaa25}
\newcommand{\putindex}[3]{\vtop{\hbox{\hspace{#3} $#1$}
            \hbox{\raise 6mm \hbox{$\scriptscriptstyle #2$}}}}

\newcommand{\gradx}[0]{\vtop{\hbox{\rm grad}
            \hbox{\raise 2.5mm \hbox{\rm \hspace{2mm} \footnotesize x}}}}

\newcommand{\grady}[0]{\vtop{\hbox{\rm grad}
            \hbox{\raise 2.5mm \hbox{\rm \hspace{2mm} \footnotesize y}}}}

\newcommand{\grad}[1]{\vtop{\hbox{\rm grad}
            \hbox{\raise 2.5mm \hbox{#1}}}}

\newcommand{\stz}{\rule{0mm}{2.3ex}}

\newcommand{\btb}{     \begin{tabbing}             }
\newcommand{\bte}{     \end{tabbing}               }

\usepackage{bm} 
\usepackage[mode=buildnew]{standalone}
\usepackage{tabularx}
\usepackage{diagbox}
\usepackage{multirow}
\usepackage[utf8]{inputenc}
\usepackage[dvipsnames]{xcolor}
\usepackage{tikz}

\makeatletter 
\newcommand\semiLarge{\@setfontsize\semiLarge{11.50}{15.00}}
\makeatother

\title{P.808 Multilingual Speech Enhancement Testing:\\
Approach and Results of URGENT 2025 Challenge}

\name{\shortstack[c]{
Marvin Sach$^{1}$, Yihui Fu$^{1}$, Kohei Saijo$^{2}$, Wangyou Zhang$^{3}$, Samuele Cornell$^{4}$, Robin Scheibler$^{5}$, Chenda Li$^{3}$, \\
Anurag Kumar$^{6}$, Wei Wang$^{3}$, Yanmin Qian$^{3}$, Shinji Watanabe$^{4}$, Tim Fingscheidt$^{1}$}
}
\address{$^{1}$	TU Braunschweig, Germany, $^{2}$ Waseda University, Japan,  $^{3}$ Shanghai Jiao Tong University, China\\
$^{4}$ Carnegie Mellon University, USA, $^{5}$ Google DeepMind, Japan, $^{6}$ Meta, USA 
}

\begin{document}

\maketitle

\begin{abstract}
In speech quality estimation for speech enhancement (SE) systems, subjective listening tests so far are considered as the gold standard. This should be even more true considering the large influx of new generative or hybrid methods into the field, revealing issues of some objective metrics. Efforts such as the Interspeech 2025 URGENT Speech Enhancement Challenge also involving non-English datasets add the aspect of multilinguality to the testing procedure. In this paper, we provide a brief recap of the \mbox{ITU-T} P.808 crowdsourced subjective listening test method. A first novel contribution is our proposed process of localizing both text and audio components of Naderi and Cutler's implementation of crowdsourced subjective absolute category rating (ACR) listening tests involving text-to-speech (TTS). Further, we provide surprising analyses of and insights into URGENT Challenge results, tackling the reliability of (P.808) ACR subjective testing as gold standard in the age of generative AI. Particularly, it seems that for generative SE methods, subjective (ACR MOS) and objective (DNSMOS, NISQA) reference-free metrics should be accompanied by objective phone fidelity metrics to reliably detect hallucinations. 
Finally, we will soon release our localization scripts and methods for easy deployment for new multilingual speech enhancement subjective evaluations according to \mbox{ITU-T} P.808.
\end{abstract}

\section{Introduction}
\label{sec:intro}

Research in the domain of speech enhancement requires frequent evaluation of a developed system's performance. Although objective metrics are very useful and usually cost-effective in determining components of speech quality, subjective listening tests performed by human listeners retain importance within the community. Especially the popularity of generative enhancement has produced new error classes such as phonetic substitutions \cite{Richter2023,Close2024}, that remain undetected by some common objective metrics \cite{Pirklbauer2023, deOliveira2023}. While employing a large set of metrics has demonstrated its usefulness in the URGENT 2024 Speech Enhancement Challenge \cite{Zhang2024}, subjective evaluation so far remains the gold standard for speech quality assessment. 

Subjective listening tests come in multiple types and implement either reference-based or reference-free testing methods according to \mbox{ITU-T} Recommendation P.800 \cite{ITU-P800}. Reference-based testing methods such as comparison category rating (CCR) tests and degradation category rating (DCR) tests, test two stimuli against each other. While CCR tests allow for direct comparison of competing systems and DCR tests are useful when measuring the degradation of a given reference, CCR tests explode in size when comparing a multitude of systems at once. DCR tests, on the other hand, are not applicable when a reference is unavailable.
A direct quality estimate without reference can be obtained from ACR subjective listening tests \cite{ITU-P800}, where listeners rate stimuli on a pre-defined scale. 

A disadvantage of subjective listening tests conducted in a controlled local environment is their association with usually major effort w.r.t.\\---among others---acquiring participants who are native speakers of the respective language, providing a controlled environment and instructing participants to ensure high quality of the obtained data.
Crowdsourcing has emerged as a noteworthy alternative to local test setups and has been verified to deliver results that are highly correlated with lab-based tests when implemented according to \mbox{ITU-T} Recommendation P.808, as demonstrated by Naderi and Cutler \cite{naderi2020,ITU-P808}. Naderi and Cutler's open-source implementation supports ACR, DCR, and CCR tests, among other methods, and leverages the Amazon MTurk platform for crowdsourcing \cite{naderi2020,naderi2021}. We consider this a reference implementation for P.808.

As part of the URGENT Speech Enhancement Challenge series, the URGENT 2024 Challenge introduced the novel aspect of generalizing the speech enhancement task to include a variety of subtasks, and provides large-scale training data combined with an extensive objective and subjective evaluation of results \cite{Zhang2024}. The second edition, URGENT 2025, extends this setup by increasing data diversity, distortion types and offering a second track with an increased amount of training data, while also introducing multilingual data to both tracks in training and evaluation \cite{Saijo2025}, as the multilingual subjective evaluation requires careful consideration. Especially, ensuring that crowdworkers are native speakers is difficult in practice. 

The explicit treatment of multilinguality is still, even in the era of emerging generative speech enhancement systems, only occasionally found in the field. Past studies claiming low language dependency report only objective metrics 
for discriminative models \cite{wang22c_interspeech}, which leaves room for investigation especially for novel generative approaches.
Listening tests in more than one language have been conducted before \cite{lechler2024,Abel2016b,Wardah2025,pia2022, Yoon2024} and have even been part of a crowdsourcing setup. Lechler and Wojcicki successfully conducted intelligibility tests \cite{lechler2024}, showing feasibility of a multilingual application of crowdsourcing to a different task using the adjacent \mbox{ITU-T} Rec.\ P.807 \cite{ITU-P807}.

While \mbox{ITU-T} Rec.\ P.808 recommends requiring native or native-level speakers of the language used for spoken material and advises an appropriate test, the P.808 reference implementation by Naderi and Cutler \cite{naderi2020} is restricted so far to English only. 

Our contributions are threefold:
\begin{itemize}
    \item \textbf{Processing approach for localization}: Building upon the English-language approach of Naderi and Cutler \cite{naderi2020}, we propose a specific technical process of localization for ACR listening tests encompassing both written and spoken instructions.
    \item \textbf{Analysis of challenge data}: We provide a thorough analysis of the multilingual objective and subjective results of the Interspeech URGENT 2025 Speech Enhancement Challenge \cite{Saijo2025}.
    \item \textbf{Open source}: Soon, we will provide our annotated scripts to the community for easy adoption, reproduction, and use for crowdsourced multilingual speech quality assessments. \phantom{https://placeholder-url.git/name/P.808-Localization}
\end{itemize}

The paper is structured as follows: In Section II, we provide an overview of crowdsourced subjective listening tests after \mbox{ITU-T} Rec.\ P.808 \cite{ITU-P808} that our work is based on \cite{naderi2020}. Furthermore, we present our various processing steps to localizing the testing procedure. Section III provides context on training and test data employed for this analysis, which took place within the scope of the URGENT 2025 Challenge. In Section IV, we provide key statistics on the listening test's participants and results of the conducted tests, and present some interesting findings. Section VI concludes the paper. 

\section{Our Testing Approach}
\begin{figure}
\centering
\resizebox{\linewidth}{!}{\includegraphics{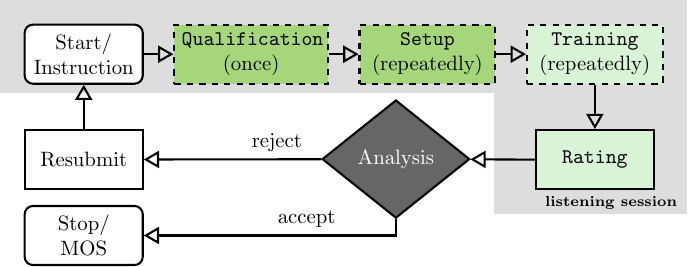}}
\caption{\textbf{P.808 listening test flow chart} and its \textbf{multiple phases}. The crowdworkers receive small parts of the dataset for a listening session (gray). While instructions and the rating prompts are presented always, a qualification test, a setup check, and training questions are only presented in the beginning and in fixed intervals, respectively. The collected ratings from all listening sessions are then analyzed for compliance with quality criteria and either used for MOS calculation or resubmitted for rating in case of rejection. From multiple listening sessions, the MOS can be calculated. The gray part of the block diagram (listening session) is visible/audible to the crowdworkers and therefore subject to our proposed localization processing (cf.\ Fig. 2) to be presented in the target language.}
\label{fig:p808-process}
\end{figure}
\subsection{P.808 Subjective Listening Tests (Recap)}
\begin{figure}[t]
    \centering
    \includegraphics[page=1, trim=4.7cm 4.7cm 14.5cm 3.3cm, clip,width=\linewidth]{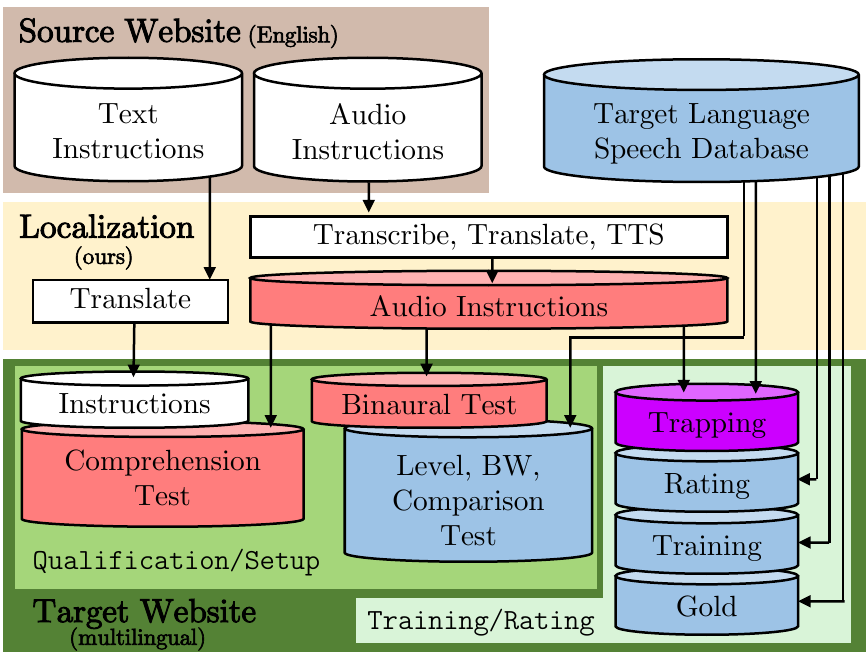} 
    \caption{\textbf{Proposed localization process}. We use a database of recorded speech (blue) in the target language to provide speech for training, gold, and P.808 trapping clips during the \texttt{training/rating} phase. Speech signals used for the level, bandwidth (BW), and comparison tests are also taken from the target language speech database. We use TTS to generate synthetic audio (red) instructions by employing Amazon's Polly with a speaker matching the target language. Trapping clips contain both synthetic and recorded audio (purple).}
    \label{fig:localization}
\end{figure}

\mbox{ITU-T} Rec.\ P.808 lays out a set of recommendations for conducting crowdsourced listening tests for speech quality evaluation \cite{ITU-P808}. A flow chart of the process is shown in \autoref{fig:p808-process}. The procedure recommended for a listening session consists of multiple components introduced as \texttt{qualification}, \texttt{setup}, \texttt{training}, and \texttt{rating} phase. The \texttt{qualification} phase is only completed once per participant and ensures participants' ability and suitable equipment, followed by a \texttt{setup} check, that tests for a quiet environment. In the \texttt{training} and \texttt{rating} phases, participants' judgement scales are calibrated, and the actual quality ratings are collected, respectively. For consistent ratings, the \texttt{setup} and \texttt{training} are repeated at fixed intervals. The gathered ratings are to be checked for reliability in an additional \texttt{analysis} step, that may discard unreliable ratings. As according to P.808 each stimulus should have a minimum number of eight ratings, rejected samples must be resubmitted to the pool of rateable clips.\\
Naderi and Cutler \cite{naderi2020} provide an English-language implementation compliant with this recommendation, leveraging the Amazon MTurk platform \cite{AmazonMTurk}, which allows the presentation of instructions and media to paid crowdworkers on a website. The components displayed to the workers are depicted in \autoref{fig:localization} in the block "Target Website".
\texttt{Qualification} consists of instructions asking for self-reported information regarding hearing ability and a comprehension test that requires speech-in-noise transcription. Additionally, a bandwidth (BW) test is performed, checking whether wideband, superwideband, and fullband recordings can be distinguished, by providing speech in bandpass-filtered white noise with increasing lower cutoff frequencies. The \texttt{setup} and environment check consist of a playback level test, a binaural listening test, asking for transcription of alternatingly left-/right-presented digits, and a comparison test, asking participants to spot a quality difference between two stimuli, induced by just noticable noise \cite{naderi2020j}. In the \texttt{training} and \texttt{rating} phase, multiple stimuli, a prompt, and multiple selectable answers are presented. The stimuli include---besides the actual \texttt{rating} and \texttt{training} clips---also gold and trapping clips. The \texttt{training} job setup follows the structure of the \texttt{rating} job and includes stimuli covering the full range of qualities expected from the rating clips and is repeated in fixed intervals.
Gold and trapping clips implement the aforementioned reliability check as gold samples are of known extremal quality that participants \textit{must} identify correctly. Trapping questions contain a spoken prompt, asking to select a specific answer. Instructions for all tests conducted and the participation rules are displayed throughout all stages of the listening session.

\subsection{Proposed P.808 Localization Process for Multilingual Tests}
\label{sec:localizedP808}
We base our proposed step-by-step localization process on the ACR implementation by Naderi and Cutler \cite{naderi2020} and pay attention to preserve and offer all functionality in the target language. An overview of localization resources and where translation occurs is shown in \autoref{fig:localization}. Using Naderi and Cutler's ACR template as the source website, we (1) \textit{manually transcribe all audio instructions} and (2) \textit{translate all text and transcribed audio instructions} to the target language. (3) For \textit{audio instructions} in the target language, we use \textit{synthetic speech} (red). To guarantee utmost quality, we ensure that all translations are performed manually by native speakers of the target language. (4) Then, all recorded utterances used in \texttt{qualification}, \texttt{setup}, and \texttt{training} are replaced by recordings in the target language (blue), which are picked from a database ($\mathcal{D}^{\mathrm{test}}_{\mathrm{non-blind}}$) disjoint from the rating clips ($\mathcal{D}^{\mathrm{test}}_{\mathrm{blind}}$). 
Unlike the original English instructions, we do not ask workers to self-report their mother tongue, but (5) \textit{only ask for fluency in the target language}, which our depicted proposed localization process ensures as \textit{we offer all instructions to workers exclusively in the target language}. Imitating the tests of the English template, (6) for the bandwidth (BW) test, we mix bandpass-filtered white Gaussian noise with lower cutoff frequencies of \mbox{4 kHz}, \mbox{9 kHz}, and \mbox{16 kHz} with clean speech. (7) For the comparison test, we mix clean speech with white Gaussian noise at SNRs of 35 dB and 45 dB.

Back to step (3): We synthesize all audio instructions using text-to-speech (TTS) employing \texttt{Amazon Polly} \cite{AmazonPolly} with a model trained for the target language. Audio instructions are employed to instruct workers during trapping questions, during the comprehension test, and during the binaural test. Trapping clips contain a short noisy segment, followed by a translation of "This is an interruption: Please select the answer $X$", where $X$ denotes the requested quality label. During the translation, special care is taken that important keywords (e.g., quality labels) are always translated consistently to the same term.

\begin{table}[t]
    \centering
    \caption{MOS and objective metrics of \textbf{all models} on $\mathcal{D}^{\mathrm{test}}_{\mathrm{blind}}$. Best is \textbf{bold}, second best \underline{underlined}.}
    \vspace{0.0cm}
    \label{tab:table1}
    
\begin{tabular}{c|c|c|c|c|c|c}
\hline\hline
\multirow{2}{*}{\!Language\!\!} & \multirow{2}{*}{\!\!MOS\!\!} & \multicolumn{5}{c}{Instrumental \stz} \\
\cline{3-7}
 & & \!\!PESQ\!\! & \!\!DNSMOS\!\! & \!\!NISQA\!\! & \!\!ESTOI\!\! & \!\!LPS\!\! \stz \\
\hline
\!EN\stz& \textbf{3.06} & 2.16 & 2.74 & 2.91 & {\textbf{0.74}} & {\textbf{0.72}} \\
\!DE & 2.97 & 2.01 & 2.86 & \underline{2.99} & 0.69 & 0.61 \\
\!ZH & {\underline{3.01}} & \underline{2.19} & \textbf{2.90} & 2.84 & 0.72 & {\underline{0.70}} \\
\!JP& {2.94} & 2.04 & \underline{2.88} & \textbf{3.08} &  {\underline{0.73}} & {0.53} \\
\hline
\!Mean\!\! \stz & {3.00} & 2.10 & 2.85 & 2.96 & 0.72 & 0.64 \\
\hline\hline
\end{tabular}

\end{table}

\section{URGENT 2025 Challenge}
The URGENT 2025 Challenge as the 2$^{\mathrm{nd}}$ edition of the URGENT Speech Enhancement Challenge series \cite{Zhang2024, Saijo2025} requires participants to develop a single model dealing with seven different types of distortions: additive noise, reverberation, clipping, bandwidth limitation, coding artifacts, packet loss and wind noise \cite{Saijo2025}. It features two tracks with \texttt{Track 1} being limited to 2.5k hours of training data and \texttt{Track 2}, which leverages 60k hours of training material. In our work, we focus on \texttt{Track 1}. Participants were asked for a self-assessment classifying their model into discriminative (D), generative (G), hybrid (H), or uncategorized (U) models.
\subsection{Datasets}
\label{subsection:datasets}
The challenge rules allow only for the usage of the official training data ($\mathcal{D}^{\mathrm{train}}$) and validation data ($\mathcal{D}^{\mathrm{val}})$. For clean speech components, the organizers provide a large-scale dataset consisting of LibriVox (DNS5) \cite{Dubey2024}, LibriTTS \cite{zen19}, VCTK \cite{VeauxVCTK}, WSJ \cite{Garofolo2007}, EARS\cite{richter2024ears}, MLS-HQ\cite{pratap20_interspeech} and CommonVoice \cite{ardila-etal-2020-common}, yielding 802 hours of purely English training data and additional 1750 hours of multilingual training data (English (EN) 550 h, German (DE) 300 h, French (FR) 300 h, Spanish (ES) 300 h, Chinese (ZH) 300 h). While data augmentation and simulation are permitted, to ensure comparability, participants must not introduce any inofficial recorded data. 
The test data is split into two parts: The non-blind test set $\mathcal{D}^\mathrm{test}_\mathrm{non-blind}$ available to participants, 
and the blind test data $\mathcal{D}^\mathrm{test}_\mathrm{blind}$, which was used for the final ranking and P.808 MOS ratings.
The blind test set is composed of the aforementioned five languages and additionally the unseen Japanese (JP) language from the databases SLR83 \cite{demirsahin2020}, SLR61 \cite{guevara-rukoz2020}, Vibravox \cite{hauret2025}, Spoken Wikipedia Corpora \cite{koehn2016}, AISHELL3 \cite{shi21c}, Hi-Fi-CAPTAIN \cite{okamoto2023}, and JVNV \cite{Xin2024}. Clean utterances were manually selected from the partly noisy corpora. Real recordings were taken from \cite{watanabe2020chime, triantafyllopoulos2024introducing, Connaeu2023, BERTIN2019,Fu2023,Li2023YODAS,ristea2024icassp}. $\mathcal{D}^{\mathrm{test}}_{\mathrm{blind}}$ consists of 150 utterances per language (50\% real, 50\% simulated), yielding 900 utterances disjoint from $\mathcal{D}^{\mathrm{train}}$, $\mathcal{D}^{\mathrm{val}}$, and $\mathcal{D}^{\mathrm{test}}_{\mathrm{non-blind}}$.

\subsection{Metrics}
The URGENT Challenge 2025 employs a multitude of objective measures to evaluate speech quality \cite{Saijo2025}. Among those are PESQ \cite{ITU-P862}, DNSMOS \cite{Reddy2021}, NISQA \cite{Mittag2021}, ESTOI \cite{jensen2016algorithm}, and the phone fidelity metric LPS (Levenshtein phone similarity = $1-\mathrm{LPD}$ (Levenshtein phone distance))\cite{Pirklbauer2023}. It measures the Levenshtein distance between the phone sequences of enhanced speech and clean speech, thereby allowing to detect hallucinations. It has already successfully been employed in the first URGENT Challenge in 2024 \cite{Zhang2024}. Reference-based objective metrics are calculated for each utterance with an available reference.
MOS (mean opinion score) is determined using the P.808 reference implementation by Naderi and Cutler \cite{naderi2020} (EN) and three of our localized variants (DE, ZH, JP), as described in Section \ref{sec:localizedP808}. We gather eight ratings per utterance for MOS calculation. All reported metrics are calculated by averaging all utterance scores of the specified subsets\footnote{Note that throughout our experimental analysis we only report on EN, DE, ZH, and JP, as the subjective results of the remaining two languages FR, ES were not yet available at time of paper submission.} of $\mathcal{D}^{\mathrm{test}}_{\mathrm{blind}}$.

\begin{table}[t]
    \centering
    \caption{MOS \textbf{by language and model type} (D: discriminative, H/G: hybrid or generative, U: uncategorized) of \textbf{all models} on $\mathcal{D}^{\mathrm{test}}_{\mathrm{blind}}$. Best is \textbf{bold}, second best \underline{underlined}.}
    \vspace{0.0cm}
    \label{tab:table2}

\begin{tabular}{c|c|c|c|c|c}
\hline\hline
\multirow{2}{*}{Model Type} & \multicolumn{4}{c|}{P.808 MOS per language \stz} & \multirow{2}{*}{MOS} \\
\cline{2-5}
 & EN\stz & DE & ZH & JP & \\
\hline
Type D \stz & \underline{3.05} & \underline{2.93} & \underline{2.97} & \underline{2.91} & {\underline{2.98}} \\
Type H/G & \textbf{3.21} & \textbf{3.07} & \textbf{3.14} & \textbf{3.03} & {\textbf{3.14}} \\
Type U & {2.86} & {2.84} & {2.85} & 2.83 & {2.85} \\
\hline
D $\cup$ H/G $\cup$ U & {3.06}\stz & {2.97} & {3.01} & 2.94 & {{3.01}} \\
\hline\hline
\end{tabular}

\end{table}

\begin{table*}[ht]
    \centering
    \caption{Multilingual MOS and LPS values and objective metrics for a \textbf{selection of models} on $\mathcal{D}^{\mathrm{test}}_{\mathrm{blind}}$ (IDs correspond to the overall model rank according to the official scoreboard \cite{Urgent2025Scoreboard, Saijo2025}. ID 10 is the baseline (BL). The best method is \textbf{bold}, the second best \underline{underlined}. Exponents denote 95\% confidence intervals.}
    \vspace{0.0cm}
    \label{tab:table3}
    
\setlength{\tabcolsep}{4pt}

\begin{tabular}{l|c|c|c|c|c|c|c|c|c|c|c|c|c|c|c}
\hline\hline
\multirow{3}{*}{\shortstack{Model\\ID}} & \multirow{3}{*}{\shortstack{Type}} & \multicolumn{4}{c}{P.808 MOS \stz} & \multicolumn{1}{c}{\diagbox[width=5.0em,height=1.4em]{}{}} & \multicolumn{4}{c}{Metrics over all languages \stz} & \multicolumn{1}{c}{\diagbox[dir=SW,width=5.0em,height=1.4em]{}{}} & \multicolumn{4}{c}{LPS by language}\\
\cline{3-6}\cline{8-11}\cline{13-16}
 & & \multirow{2}{*}{EN} & \multirow{2}{*}{DE} & \multirow{2}{*}{ZH} & \multirow{2}{*}{JP} & \multirow{2}{*}{{MOS}} & \multirow{2}{*}{PESQ} & \multirow{2}{*}{DNSMOS} & \multirow{2}{*}{NISQA} & \multirow{2}{*}{ESTOI} & \multirow{2}{*}{LPS} & \multirow{2}{*}{EN} & \multirow{2}{*}{DE} & \multirow{2}{*}{ZH} & \multirow{2}{*}{JP}\\
 & & & & & & & & & & & & & & &\\
\hline
Noisy\stz & - & 2.13 & 1.92 & 1.84 & 1.73 & 1.91$^{\pm 0.05}$ & 1.31 & 1.90 & 1.58 & 0.58 & 0.49$^{\pm 0.04}$ & 0.55 & 0.47 & 0.59 & 0.36 \\
1 & D & 3.24 & 3.06 & 3.16 & 3.00 & 3.12$^{\pm 0.06}$ & \textbf{2.61} & 2.90 & 3.23 & \textbf{0.82} & \textbf{0.73}$^{\pm 0.03}$ & \textbf{0.81} & \textbf{0.71} & \textbf{0.78} & \textbf{0.60} \\
2 & H & {3.32} & \textbf{3.36} & {3.31} & \underline{3.26} & 3.31$^{\pm 0.06}$ & \underline{2.44} & 2.92 & 3.24 & 0.78 & 0.70$^{\pm 0.03}$ & {0.78} & 0.65 & \underline{0.76} & \underline{0.59} \\
3 & H & \underline{3.44} & \underline{3.33} & \textbf{3.39} & \textbf{3.39} & \textbf{3.39}$^{\pm 0.05}$ & 2.41 & 2.95 & 3.24 & 0.78 & 0.70$^{\pm 0.03}$ & 0.78 & \underline{0.68} & 0.74 & {0.58} \\
4 & H & 3.28 & 3.02 & 3.20 & 2.96 & 3.12$^{\pm 0.06}$ & \underline{2.44} & 2.81 & 3.03 & 0.79 & 0.70$^{\pm 0.03}$ & 0.78 & 0.67 & \underline{0.76} & \underline{0.59} \\
5 & H & 3.04 & 2.79 & 2.84 & 2.71 & 2.85$^{\pm 0.06}$ & \underline{2.44} & 2.84 & 2.92 & \underline{0.80} & \underline{0.71}$^{\pm 0.03}$ & \underline{0.79} & \underline{0.68} & \underline{0.76} & \underline{0.59} \\
6 & H & 3.21 & 3.13 & 3.21 & 3.14 & 3.17$^{\pm 0.06}$ & 2.23 & 2.92 & 3.29 & 0.77 & 0.69$^{\pm 0.03}$ & {0.78} & 0.64 & 0.75 & \underline{0.59} \\
7 & H & 3.19 & 2.95 & 3.07 & 3.01 & 3.06$^{\pm 0.06}$ & 2.43 & 2.87 & 3.08 & 0.78 & 0.67$^{\pm 0.03}$ & 0.75 & 0.64 & 0.73 & 0.57 \\
8 & D & 3.17 & 3.08 & 3.04 & 3.03 & 3.08$^{\pm 0.06}$ & 2.23 & 2.93 & 3.14 & 0.76 & 0.66$^{\pm 0.03}$ & 0.74 & 0.60 & 0.73 & 0.55 \\
9 & H & 3.17 & 3.14 & 3.17 & 3.08 & 3.14$^{\pm 0.06}$ & 1.95 & \underline{3.08} & \underline{3.70} & 0.73 & 0.65$^{\pm 0.03}$ & 0.72 & 0.60 & 0.72 & 0.54 \\
10 (BL) & D & 2.96 & 2.80 & 2.89 & 2.88 & 2.88$^{\pm 0.06}$& 2.21 & 2.86 & 2.76 & 0.76 & 0.66$^{\pm 0.03}$ & 0.74 & 0.61 & 0.73 & 0.56 \\
11 & U & 2.94 & 2.83 & 2.77 & 2.86 & 2.85$^{\pm 0.06}$ & 2.01 & 2.87 & 3.04 & 0.74 & 0.63$^{\pm 0.03}$ & 0.73 & 0.59 & 0.70 & 0.48 \\
12 & D & 2.82 & 2.78 & 2.78 & 2.75 & 2.78$^{\pm 0.07}$ & 2.37 & 2.80 & 2.76 & 0.74 & 0.64$^{\pm 0.03}$ & 0.70 & 0.60 & 0.71 & 0.53 \\
13 & G & \textbf{3.69} & 3.23 & \underline{3.33} & 3.11 & \underline{3.34}$^{\pm 0.05}$ & 1.33 & \textbf{3.13} & \textbf{3.79} & 0.53 & 0.58$^{\pm 0.03}$ & 0.71 & 0.61 & 0.56 & 0.44 \\
\hline\hline
\end{tabular}

\end{table*}

\section{Experimental Analysis}

\subsection{Subject Characterization}
The crowdworkers employed for subjective testing are fully characterized by their compliance with the requirements checked during \texttt{qualification} and \texttt{setup} phase. We observed, however, a significantly lower acceptance rate w.r.t. our reliability check \texttt{analysis} for non-English listening tests. While English listening tests had an acceptance rate of 82.3\%, German, Chinese, and Japanese listening tests showed acceptance rates of only 62.8\% and 41.1\%, and 28.9\% respectively, during the first stage of testing. Excluding crowdworkers with high rejection rates from re-rating the rejected utterances, this rate was slightly improved after resubmission (cf.\ Fig.\ \ref{fig:p808-process}).

\subsection{Subjective and Objective Results}
The URGENT 2025 Challenge results of the multilingual subjective listening tests along with some objective metrics are displayed in Tables \ref{tab:table1}, \ref{tab:table2}, \ref{tab:table3}. 

Table \ref{tab:table1} shows the MOS values and select objective metrics across all participating models (including the baseline) on four languages which form subsets of $\mathcal{D}^{\mathrm{test}}_{\mathrm{blind}}$. We observe high(est) scores for MOS and intelligibility / phone fidelity metrics (ESTOI, LPS) for the English subset of $\mathcal{D}^{\mathrm{test}}_{\mathrm{blind}}$. This may be partially attributed to the imbalance of training data, which features an additional 802 hours of English speech besides the English speech included in the 1750 hours of multilingual data. The unseen language Japanese {shows decreased subjective quality} accompanied by a dramatic drop of phonetic fidelity (LPS $=0.53$), which is not reflected by the ESTOI metric. Interestingly, we observe the same rank order among languages for MOS and LPS. While the unseen Japanese language shows overall best reference-free metric results (DNSMOS 2$^{\mathrm{nd}}$ rank, NISQA 1$^\mathrm{st}$ rank), its overall MOS score is lowest among the four languages.

Table \ref{tab:table2} resolves MOS data for each language a step further by differentiating between model categories (D, H/G mixed, and U).
We observe that the class of H/G systems performs best across all languages in subjective evaluation. The MOS rankings of model types is consistent in each particular language, led by type H/G, followed by type D, and type U being last. This is in line with the overall ranking of models on the official scoreboard \cite{Urgent2025Scoreboard}, where type H/G models are grouped at the top, type U models are grouped at the bottom and type D models are scattered in the top half.

Table \ref{tab:table3} shows subjective and objective metrics by model (type), with MOS and LPS also being analyzed in individual languages. Each row shows a different model ID corresponding to the ranking in the official scoreboard \cite{Urgent2025Scoreboard}.

We see that the discriminative model \#1 scores highest in all reference-based objective metrics PESQ (significantly highest $2.61$), ESTOI ($0.82$), and LPS ($0.73$), however, falls behind the top MOS-scoring model \#3 consistently in each investigated language. 

The generative model \#13 shows by far poorest overall intelligibility and phone fidelity (ESTOI $= 0.53$, LPS $= 0.58$) of all reported methods in Table \ref{tab:table3}. This is in significant contrast to the top scores of method \#13 in the reference-free objective metrics DNSMOS ($3.13$) and NISQA ($3.79$). We can conclude from this example that DNSMOS  and NISQA  (dependent on the speech enhancement technology) may ignore the aspect of intelligibility and phone fidelity, whereas former literature states that speech quality includes intelligibility to some extent; both correlate until intelligibility is saturated \cite{Arehart2022, Schiffner2014}. {Chiang et al.\ \cite{Chiang2023} find that subjective quality and subjective intelligibility have a Pearson correlation coefficient exceeding even the correlation between subjective quality and PESQ, and between subjective quality and DNSMOS. Informal inspection of method \#13 in DE and in EN confirms a high degree of hallucination. To summarize, we can state that \textit{reference-free objective metrics may very much ignore aspects such as intelligibility and phone fidelity (or hallucination).}

It becomes even worse: Also the high MOS score ($3.34$) of method \#13 is in surprising contrast to the low intelligibility and phone fidelity scores of method \#13. We could state: "It sounds like English [or any other language of the test], but it is not." To the best of our knowledge, for the first time,
{\it the irritating observation is reported that (P.808) ACR listening tests (which are by nature also reference-free just as DNSMOS and NISQA) may simply ignore aspects such as intelligibility and phone fidelity (or hallucination).} 
In the age of generative AI, can subjective ACR listening tests then still serve as gold standard? We tend to strongly recommend to ensure native listeners when judging generative speech enhancement methods, which is, however, difficult in the ITU-T P.808 crowdsourcing-based testing protocol.
\textit{Our recommendation is that MOS results from P.808 ACR listening tests shall be accompanied by metrics that test for intelligibility (e.g., ESTOI), or for phone fidelity (e.g., LPS).} This justifies the URGENT Challenges' approach of a ranking obtained from average rankings in many objective metrics and MOS \cite{Zhang2024,Saijo2025}. High reference-free subjective or objective metrics coming along with a drop in ESTOI or particularly in LPS may signify a high degree in hallucinations.

\section{Conclusions}
In this work, we proposed a method localizing ITU-T P.808 subjective speech quality tests for multilingual testing. We provide detailed instructions and code for adoption by the community. The results of our multilingual tests have been presented in the context of the URGENT 2025 Speech Enhancement Challenge. 
Our analysis shows that \textit{both objective and subjective reference-free} speech quality evaluations may provide questionable results when being applied to generative methods, as they have been shown (in some cases) to be insensitive to hallucinations. Future research on subjective evaluation of generative speech enhancement methods may address the aspect of native listening subjects a bit deeper, as well as the potential of comparative category rating (CCR) tests. For the moment, we must be aware that good reference-free subjective or objective metrics coming along with a drop in LPS may signify a high degree in hallucinations.

\bibliographystyle{IEEEtran}
\bibliography{ifn_spaml_bibliography}

\end{document}